# Alfvén Wave Driven High Frequency Waves in the Solar Atmosphere: Implications for Ion Heating

Edisher Kh. Kaghashvili

Live Science and Technologies, LLC.

1R Snow Circle, Nashua, NH 03062

ekaghash@lstnh.com

**ABSTRACT**

This work is an extension of *Kaghashvili* [1999] where ion-cyclotron wave dissipation channel for Alfvén waves was discussed. While our earlier study dealt with the mode coupling in the commonly discussed sense, here we study changes in the initial waveform due to interaction of the initial driver Alfvén wave and the plasma inhomogeneity, which are implicitly present in the equations, but were not elaborated in *Kaghashvili* [1999]. Using a cold plasma approximation, we show how high frequency waves (higher than the initial driver Alfvén wave frequency) are generated in the inhomogeneous solar plasma flow. The generation of the high frequency forward and backward propagating modified fast magnetosonic/whistler waves as well as the generation of the driven Alfvén waves is discussed in the solar atmosphere. The generated high frequency waves have a shorter dissipation timescale, and they can also resonant interact with particles using both the normal cyclotron and anomalous cyclotron interaction channels. These waves can also be important in space, Earth's magnetosphere, the ionosphere, and laboratory plasma processes.

**Introduction**

In 1965, when comparing the theoretical predictions with the existing observational data, Parker [1965] concluded that the escaping solar wind plasma had to be actively heated in the solar corona and for some considerable distance when traveling from the Sun towards the Earth and beyond. Since then, questions like "What heats the solar corona?" or "What mechanism energizes the solar wind plasma?" motivated a number of studies that later were proved instrumental in furthering our understanding of the solar wind origin and its dynamics [e.g., see reviews by *Hollweg*, 1990; *Hollweg and Isenberg*, 2002; *Cranmer*, 2002; *Ofman*, 2005; *Hollweg*, 2006, 2008; *Cranmer*, 2009]. One of those studies was about possible effects of the ion-cyclotron waves on the escaping solar plasma particles.

Cyclotron interactions between waves and particles play an important role in the plasma heating and energization of the plasma particle populations [e.g., *Lee and Ip*, 1987; *Galeev et al.*, 1991; *Isenberg*, 2001; *Hollweg and Isenberg*, 2002; *Isenberg and Vasquez*, 2007; *Yoon and Seough*, 2012; *Galinsky and Shevchenko*, 2013; *Kasper et al.*, 2013]. Due to the effective interaction of near cyclotron frequency waves and particles and the high temporal sampling required to detect such high frequency waves, a majority of works that deal with such processes in space, solar, magnetosphere and laboratory plasmas, are divided into two groups: (a) studies that assume the existence of such waves and investigate their effects on particles, and (b) studies that deal with the generation of such waves in the respective plasma environment, i.e. how these waves are generated and how their spectrum can be maintained as well.

By postulating that the Sun launched low-frequency waves would transfer and generate the high frequency ion-cyclotron waves through the turbulent cascade, *Hollweg* [1986; see also *Hollweg and Johnson*; 1988; *Isenberg*, 1990] included the effects of the ion-cyclotron waves on the solar plasma particles in the solar wind model. It was shown that ion cyclotron resonance would heat and accelerate the protons and other solar wind ion species. This prediction, which was revolutionary at that time, was dismissed due to a lack of data, but it was later confirmed by the Ultraviolet Coronagraph Spectrometer (UVCS/SOHO) observations that indicated preferential heavy ion heating and acceleration close to the Sun [*Kohl et al.*, 1998]. Detailed nature of the ion-cyclotron wave interaction with particles in the solar wind [e.g., *Markovskii et al.*, 2006; *Markovskii et al.*, 2010; *Kasper et al.*, 2013] and the modeling the large scale effects of such interactions [e.g., *Marsch*, 1999; *Cranmer et al.*, 1999; *Hu et al.*, 2000; *Cranmer et al.*, 2004] have still remain an active area of research.

While the efficient dissipation of the waves resonant with ion-cyclotron motions about the coronal magnetic field lines can explain the observations of ion species, it is no less important to understand what actually generates such near ion-cyclotron frequency waves in the solar atmosphere, and what mechanism maintains their spectrum long enough to produce the extended heating and acceleration of the solar wind. The first possible source was suggested to be Alfvén wave saturation followed by a turbulent cascade from low to high, near ion-cyclotron frequency waves [*Isenberg and Hollweg*, 1983]. *Axford and McKenzie* [1992] proposed that reconnection processes associated with coronal flare activities can generate high-frequency waves. The later suggestion is often referenced in scientific works where possible effects of the ion-cyclotron waves are modeled in the solar chromosphere, solar corona and solar wind [e.g., *Axford et al.*, 1999; *Markovskii and Hollweg*, 2004; *Marsch*, 2006; *Hansteen and Velli*, 2012; *Cranmer*, 2012;

*Xiong and Li*, 2012; *Galinsky and Shevchenko*, 2013]. Although, as mentioned above, the dissipation rate of the near ion-cyclotron frequency waves is strong. As a result, these waves are not capable of propagating long distances to produce an extended heating. Thus there is a need to understand what produces and maintains the spectrum of such waves in the solar corona and solar wind.

More than a decade ago, *Kaghashvili* [1999; see also *Kaghashvili*, 2002; *Chen*, 2005] considered the mechanism of high frequency wave generation from the "original" wave spectrum of the photosphere. The mechanism was based on the gradual changes in the initial wave vector that allow the transformation of one mode to a different mode in the flowing magnetized plasma [*Chagelishvili et al.*, 1996]. Here, we extend *Kaghashvili* [1999] work using the driven wave formalism [*Kaghashvili*, 2007; 2013; *Hollweg and Kaghashvili*, 2009; *Hollweg et al.*, 2013]. We consider the evolution of the initial Alfvén wave outside the coupling region (the coupling region is defined as the frequency region where frequencies of two interacting modes are nearly the same) and show that a more complex interaction is implicitly present in the equations, which was not elaborated in *Kaghashvili* [1999]. We show how an initial driver Alfvén wave can generate high-frequency driven waves in the solar atmosphere. (By high frequency we mean the frequency higher than the initial driver wave frequency, which can range from the MHD regime till frequencies for which the non-relativistic kinetic description of the electron-proton plasma given below is valid.) While the high frequency driven wave physics described here can equally be applicable to all plasma environments, like the space plasma, the Earth's magnetosphere, the ionosphere and laboratory plasma processes where kinetic waves play an important role [e.g., *Gekelman et al.*, 1997; *Amatucci et al.* 1998; *Peñano et al.*, 1998; *Amatucci*, 1999; *Génot et al.*,

1999; *Koepke et al.*, 1999; *Kaneko et al.*, 2002; *Koepke et al.*, 2003; *Chaston et al.*, 2004; *Koepke and Reynolds*, 2007; *Bering et al.*, 2008; *Koepke*, 2008; *Lysak and Song*, 2008; *Shukla and Saleem*, 2008; *Birn et al.*, 2010; *Bering et al.*, 2010; *Blackwell et al.*, 2010; *Brunner et al.*, 2013; *DuBois et al.*, 2013; *Perron et al.*, 2013], here our focus is on solar coronal and solar wind plasmas.

**Physical Description**

Consider a simple case of the homogeneous background magnetic field aligned flow in the electron-proton plasma. The x-axis is parallel to the homogeneous background magnetic field $\mathbf{B}_0 = (B_0, 0, 0)$, where $B_0 = const.$ The background flow for both electrons and protons is $V_{mean} + V_0$ along the magnetic field, where $V_{mean}$ is arbitrary $V_0$ characterizes the shearing of the flow along the y-axis. The coordinate system is fixed relative to the spatially constant mean flow, $V_{mean} = const.$, and it is in this frame where the background flow with the linear cross-field shear is given by

$$\mathbf{V}_0 \equiv \mathbf{V}_{0\{e,p\}} = (S_y y,\ 0,\ 0), \tag{1}$$

where $S_y = const.$ is the shear parameter, and indexes $\{e, p\}$ stand for electrons and protons, respectively.

Following the linear wave theory, all variables can be decomposed into mean and perturbed components. In the linearized equations, we follow standard steps to simplify two-fluid equations by introducing new one-fluid variables: $\rho = m_e n_e + m_p n_p$, $\mathbf{v} = (m_e \mathbf{v}_e + m_p \mathbf{v}_p)/(m_e + m_p)$, and make some general assumptions about the property of the background plasma and linear waves:

(a) the process in non-relativistic, which assumes that $\partial_t \mathbf{E} \approx 0$ and (b) plasma is quasi-neutral, i.e. $\rho_q = e(n_p - n_e) = 0$. As a result, the final set of equations is given by:

$$\frac{\partial \rho}{\partial t} + (\mathbf{V}_0 \cdot \nabla)\rho + \rho_0 (\nabla \cdot \mathbf{v}) = 0 \tag{2}$$

$$\frac{\partial \mathbf{v}}{\partial t} + (\mathbf{V}_0 \cdot \nabla)\mathbf{v} + (\mathbf{v} \cdot \nabla)\mathbf{V}_0 = -\frac{C_{eff}^2}{\rho_0}\nabla \rho + \frac{1}{4\pi\rho_0}[\nabla \times \mathbf{b}] \times \mathbf{B}_0 \tag{3}$$

$$\frac{\partial}{\partial t}\left[(1+\lambda^2 k^2)\mathbf{b}\right] + (\mathbf{V}_0 \cdot \nabla)\left[(1+\lambda^2 k^2)\mathbf{b}\right] = \left([(1+\lambda^2 k^2)\mathbf{b}] \cdot \nabla\right)\mathbf{V}_0 + (\mathbf{B}_0 \cdot \nabla)\mathbf{v} - \mathbf{B}_0(\nabla \cdot \mathbf{v})$$

$$+ \lambda^2 ([\nabla \times \mathbf{V}_0] \cdot \nabla)[\nabla \times \mathbf{b}] - \frac{c(m_p - m_e)}{4\pi e\rho_0}(\mathbf{B}_0 \cdot \nabla)[\nabla \times \mathbf{b}] \tag{4}$$

$$\nabla \cdot \mathbf{b} = 0 \tag{5}$$

where $\lambda = (m_\mu c^2 / 4\pi e^2 n_0)^{1/2}$ is the collisionless skin depth, $m_\mu = m_e m_p /(m_e + m_p) \cong m_e$ is a normalized mass, $C_{eff}^2 = k_B (n_e \gamma_e T_e + n_p \gamma_p T_p)/(m_e n_e + m_p n_p)$ is an effective sound speed taken to be constant. These equations have been used in the velocity-shear wave coupling studies previously [*Kaghashvili*, 1999, 2002; *Chen*, 2005]. In MHD limit, i.e. when $\lambda \cong 0$ and the characteristic frequencies of the system are much smaller than $\Omega_p$ (the proton gyro-frequency), the system reduces to the more familiar form used in standard MHD equations.

For the sake of the simplicity, we consider a cold plasma case $C_{eff} = 0$, i.e. effects of the plasma thermal pressure will be ignored. This is a restriction of our current analysis, and this is the approximation that is commonly used when studying the wave phenomenon in the magnetically dominated plasma [e.g., *Kaghashvili*, 1999; *Hollweg and Kaghashvili*, 2012]. In the absence of the background flow, the above equations lead to the following dispersion equation:

$$\omega^4 - \left[\frac{(k_x^2+k^2)v_a^2}{1+k^2\lambda^2} + \frac{k^2v_a^2}{(1+k^2\lambda^2)^2}\frac{k_x^2v_a^2}{\Omega_\mu^2}\right]\omega^2 + \frac{k_x^2k^2v_a^4}{(1+k^2\lambda^2)^2} = 0 \tag{6}$$

which gives two different modes, modified Alfvén and fast magnetosonic modes:

$$\omega^2_{\{Alf, fast\}} = \frac{1}{2}\left[\frac{(k_x^2+k^2)v_a^2}{1+k^2\lambda^2} + \frac{k^2v_a^2}{(1+k^2\lambda^2)^2}\frac{k_x^2v_a^2}{\Omega_\mu^2}\right]$$

$$\mp \frac{1}{2}\sqrt{\left[\frac{(k-k_x)^2v_a^2}{1+k^2\lambda^2} + \frac{k^2v_a^2}{(1+k^2\lambda^2)^2}\frac{k_x^2v_a^2}{\Omega_\mu^2}\right] \times \left[\frac{(k+k_x)^2v_a^2}{1+k^2\lambda^2} + \frac{k^2v_a^2}{(1+k^2\lambda^2)^2}\frac{k_x^2v_a^2}{\Omega_\mu^2}\right]} \tag{7}$$

where $k_x$ and $k$ are the initial wave along the background magnetic field direction and it's magnitude, respectively. $v_a$ is an Alfvén speed, $\Omega_\mu = \frac{\Omega_e \Omega_p}{\Omega_e + \Omega_p} \approx \Omega_p$ is an effective cyclotron frequency, and $\Omega_e$ and $\Omega_p$ are the electron gyro-frequency and proton gyro-frequency, respectively. It is obvious from the Eq. (7) that these two frequencies always differ from each other. In the kinetic regime, $\omega_{Alf}$ and $\omega_{fast}$ are the frequencies of the kinetic Alfvén wave and the whistler wave. Figures 1 shows the $\omega$ vs $kv_a/\Omega_p$ dependence plots for the modified/kinetic Alfvén and fast magnetosonic/whistler waves given by Eq. (7) for the $[10^{-2}, 10^3]$ range of $kv_a/\Omega_p$ value. Three representative cases are chosen depending on the initial wave propagation angle between $\mathbf{k}$ and $\mathbf{B}_0$. As expected, the separation between these two frequencies depends on the initial Alfvén wave propagation angle and increases with $kv_a/\Omega_p$ value.

## High-Frequency Driven Waves

Consider an arbitrary polarized initial driver Alfvén wave propagating in the above described plasma environment along $\mathbf{B}_0$. The linearized equations (2)-(5) are spatially inhomogeneous due to the inhomogeneous background flow. To analyze the temporal evolution of the initial driver Alfvén wave, we change variables from the laboratory to convecting Lagrangian coordinates [e.g., *Goldreich and Lynden-Bell*, 1965; *Phillips*, 1966; *Craik and Criminale*, 1986; *Chagelishvili et al.*, 1996; *Kaghashvili*, 1999, 2007; *Camporeale*, 2012]: $r \rightarrow r - \mathbf{V}_0 dt$ and $t \rightarrow t$, which converts the spatial inhomogeneity associated with the velocity shear in Eq. (2)-(5) into a temporal one. Afterwards, the spatial Fourier expansion of the fluctuating quantities is performed, which vary as $\sim \exp(ik_x x + iK_y y + ik_z z)$, where $k_x$ and $k_z$ are constants and the wave vector along the shear changes with time as expected from the geometrical optics: $K_y(t) = k_y(t=0) - k_x S_y t$. Next we objective is to study the temporal behavior of the spatial Fourier harmonics in time.

If using the commonly accepted mode coupling physics [e.g., *Melrose*, 1977a-b, *Wentzel*, 1989; *Chagelishvili et al.*, 1996, 1997; *Poedts et al.*, 1998; *Mahajan and Rogava*, 1999; *Rogava et al.*, 2000; *Gogoberidze et al.*, 2004; *Webb et al.*, 2005a-b; *Shergelashvili et al.*, 2006; *Rogava and Gogoberidze*, 2005; *Webb et al.*, 2007; *Gogoberidze et al.*, 2007] as it was done in *Kaghashvili* [1999], two frequencies given by Eq. (7) need to be nearly in phase to each other in order to transform one MHD mode into another efficiently. If two frequencies are far from each other and one seeks for wave solutions with amplitudes explicitly independent of the inhomogeneities in the background plasma (as it is commonly done in the plasma literature), then the Alfvén wave will propagate unaffected. Using the driven wave formalism, we challenged that view. This new treatment does not change the accepted view about the coupling, i.e. the coupling is the strongest when the frequencies of the different modes are close [*Hollweg et al.*, 2013], but it reveals the more complex nature of wave interactions even when the frequencies of the modes are far from each other.

To demonstrate this, we consider a short-term evolution of the initial Alfvén wave when no changes in the initial Alfvén wave are expected according to WKB formalism. To avoid the coupling between the natural modes, we will consider the initial launched Alfvén wave is oblique relative to $\mathbf{B}_0$. Following the driven wave formalism [e.g., *Kaghashvili*, 2013], our goal is to obtain the analytical solutions for the initial driver wave and flow inhomogeneity interaction driven waves. For the short-time evolution, the typical step is to take $K_y \cong k_{y0} = const.$ (This assumption is less restrictive when the shear parameter is small and the time-dependent wave vector along the velocity shear does not change appreciably from its initial value during the time considered.) More accurate analytical solutions for the driven waves can be obtained when the "freezing" procedure on the wave vector is done in the system of the second-order differential equations. In this case, our short-time evolution assumption states that during the time-scale of interest no significant changes occur in the natural frequencies of the plasma given by Eq. (7). Before implementing above described procedure, Equations (2)-(5) can be written as:

$$\frac{d^2 v_y}{dt^2} + \frac{(k_x^2 + K_y^2) v_a^2}{1 + \lambda^2 k^2} v_y + \frac{K_y k_z v_a^2}{1 + \lambda^2 k^2} v_z + i \frac{k_x v_a}{\Omega_\mu} \frac{k^2 v_a^2}{1 + \lambda^2 k^2} b_z = f_{VY}$$

$$\frac{d^2 v_z}{dt^2} + \frac{K_y k_z v_a^2}{1 + \lambda^2 k^2} v_y + \frac{(k_x^2 + k_z^2) v_a^2}{1 + \lambda^2 k^2} v_z - i \frac{k_x v_a}{\Omega_\mu} \frac{k^2 v_a^2}{1 + \lambda^2 k^2} b_y = f_{VZ}$$

$$\frac{d^2 b_y}{dt^2} + \left[ \frac{(k_x^2 + K_y^2) v_a^2}{1 + \lambda^2 k^2} + \frac{k_x^2 v_a^2}{\Omega_\mu^2} \frac{k^2 v_a^2}{(1 + \lambda^2 k^2)^2} \right] b_y + \frac{K_y k_z v_a^2}{1 + \lambda^2 k^2} b_z + i \frac{k_x v_a}{\Omega_\mu} \left[ \frac{K_y k_z v_a^2}{1 + \lambda^2 k^2} v_y + \frac{(k_x^2 + k_z^2) v_a^2}{1 + \lambda^2 k^2} v_z \right] = f_{BY}$$

$$\frac{d^2 b_z}{dt^2} + \frac{K_y k_z v_a^2}{1 + \lambda^2 k^2} b_y + \left[ \frac{(k_x^2 + k_z^2) v_a^2}{1 + \lambda^2 k^2} + \frac{k_x^2 v_a^2}{\Omega_\mu^2} \frac{k^2 v_a^2}{(1 + \lambda^2 k^2)^2} \right] b_z - i \frac{k_x v_a}{\Omega_\mu} \left[ \frac{(k_x^2 + K_y^2) v_a^2}{1 + \lambda^2 k^2} v_y + \frac{K_y k_z v_a^2}{1 + \lambda^2 k^2} v_z \right] = f_{BZ}$$

where $f$'s on the right hand side are all shear parameter dependent "forcing" terms that will modify the initial waveform and excite the driven waves. As an example, the forcing term for the y-component of the velocity is given by: $f_{VY} = -iS_y \left( \frac{2(1+\lambda^2 k_z^2)}{1+\lambda^2 k^2} K_y b_z + \frac{1+2\lambda^2(1+k_z^2)}{1+\lambda^2 k^2} k_z b_z \right)$.

It is straightforward to write all other "forcing" terms, which are not provided here. Without the $f$ terms, one can easily verify that the natural frequencies of the second order differential equations are given by Eq. (7).

Afterwards, we take $K_y \cong k_{y0}$ in above equations and obtain solutions for all four variables using the Laplace transform. Since the shear parameter $S_y$ is small, one can expand the solution in the Laplace space as the series in terms of this parameter. As an example, the time-and-space dependent solutions of the y-component perpendicular velocity fluctuations are given by:

$$\frac{v_y}{v_{y0}} = \sum_{\pm} A_{\{a,f\}0\pm} \exp(\pm i\omega_{\{a,f\}0} t) + A_{\{a,f\}1\pm} \exp(\pm i\omega_{\{a,f\}1} t) + A_{\{a,f\}2\pm} \exp(\pm i\omega_{\{a,f\}2} t) + \ldots \qquad (8)$$

where $A_{\{a,f\}n\pm} \propto S_y^n$ are amplitudes of the respective waves in the forward and backward directions. $n$ donates the order of approximation in terms of the shear parameter, and $\omega_{\{a,f\}n}$ are the general forms of the oscillating frequencies of n-th order terms. The shear parameter independent wave amplitudes, $A_{\{a,f\}0\pm}$, represent the initial driver Alfvén wave. All other components in Eq. (8) have shear parameter dependent amplitudes, and represent the driven waves excited in the system as a result of the initial driver wave and background velocity shear flow coupling. The solutions of all other linear fluctuating components will have the similar form.

Since $K_y \cong k_{y0}$ approximation used in deriving the governing equations above, the driven wave solutions are valid for a limited time (typically a few Alfvén timescales). The oscillating frequencies in Eq. (8) are given by:

$$\omega_{\{a,f\}n} \cong \omega_{\{Alf,fast\}} + \omega_{\{a,f\}n1}S_y + \omega_{\{a,f\}n2}S_y^2 + O(S_y^3), \tag{9}$$

where the first terms (shear parameter independent) on the right hand side are the frequencies of the modified/kinetic Alfvén and fast magnetosonic/whistler waves given by Eq. (7). The specific expression of the frequency correction terms, $\omega_{\{a,f\}nk}$, depend on the approximation method used when solving the formal solutions in Laplace space [e.g., see *Kaghashvili*, 2013 for simple treatment, and *Hollweg et al*. 2013 for more elaborate treatment]. The analytical expressions of the driven wave amplitudes and frequency corrections are dependent on the initial wave characteristics (i.e., amplitude, polarization, wave vectors, etc.), plasma characteristics and the shear parameter.

Finally, let us estimate a few cases of the Alfvén wave driven high frequency waves. The plasma parameters and initial arbitrary polarized driver Alfven parameters are taken to satisfy the following condition: $k_x v_a / \Omega_p = 0.01$. Figure 1 shows that when the propagation angle, theta = $75^0$ degrees, the frequency of the initial Alfvén wave and the frequencies of two, forward and backward propagating, driven Alfvén waves are $\omega_{Alf}/\Omega_p = 0.01$, and the frequency of the two backward and forward propagating driven fast magnetosonic waves is $\omega_{fast}/\Omega_p = 0.0386$. For theta=$15^0$, the separation between these two frequencies is modest: $\omega_{Alf}/\Omega_p = 0.01$ and $\omega_{fast}/\Omega_p = 0.0104$.

**Discussion and Summary**

To summarize, we showed how the initial driver Alfvén wave can generate higher frequency driven fast/whistler waves in the solar atmosphere. We extended our earlier work [Kaghashvili, 1999] to show that Alfvén (or cyclotron/ion-cyclotron wave in high frequency regime) drives up higher frequency forward and backward fast/whistler waves even when frequencies given by Eq. (7) are not in phase. In previous works [e.g., Hollweg and Kaghashvili, 2012; Hollweg et al., 2013; Kaghashvili, 2013], we showed how Alfvén waves in a shear flow can drive up the other two MHD modes, viz. the fast and slow modes. For oblique propagation the resulting fast mode will have a higher frequency than that of the original Alfvén wave. Here we extended analysis out of the MHD regime, but for a cold plasma in which only cyclotron/ion-cyclotron and fast/whistler waves will be present. Since the oblique fast/whistler is not purely circularly polarized in the electron-resonant sense, protons and ions can undergo a cyclotron-resonant interaction with part of the fast/whistler power. We suggested that this process could contribute to heating and acceleration of protons and ions in the solar wind and corona.

For vary low frequency MHD waves, implications for the proton heating in the solar wind and corona can be modest, for the following reasons:

1. For the low frequency modes, even for waves initially propagating at $80^0$ degrees with respect to the background magnetic field, Figure 1 shows that a fast/whistler wave with $\omega = \Omega_p$ requires an initial Cyclotron/ion-cyclotron wave with $\omega = 0.176\Omega_p$. Smaller wave propagation angles or lower Cyclotron/ion-cyclotron frequencies produce lower frequency fast/whistlers. For reasonable propagation directions, a cyclotron-resonant wave cannot be generated from an initial Alfvén wave with $\omega \ll \Omega_p$, but only from an initial wave which is already out of the MHD regime;

2. With large propagation angles, the fast/whistler wave amplitudes are very weak. For example, using the MHD results as a guide, Equations (18)-(19) in Hollweg and Kaghashvili [2012] show that the amplitude of the fast wave is approximately proportional to $1/k$ when $k \gg k_x$. Nevertheless, despite the small amplitudes for $k \gg k_x$, the process continuously will draw the energy from plasma inhomogeneity and deposit it into highly dissipative fast/whistlers waves.

For an initial, arbitrary polarized Alfvén wave driver, we showed how one can obtain driven wave solutions that describe the evolution of driven waves and these solutions are valid for the initial kinetic Alfvén waves as well. Analytical solutions for a specific driver wave, in principal, would allow us to estimate: (a) the relative contribution of all possible driven waves, (b) the likelihood that the corresponding component of the driven wave fluctuations can be observationally detected, and (c) the possible effects on particles [e.g., *Khazanov et al.*, 1999; *Khazanov and Gamayunov*, 2007; *Kaghashvili*, 2012; *Yoon et al.*, 2012; *Verscharen et al.*, 2013; *Isenberg et al.*, 2013; *Khazanov and Krivorutsky*, 2013].

As was shown, in the cold plasma approximation, the arbitrary polarized initial driver wave with either frequency given by Eq. (7), in general, will generate four driven waves (waves with two characteristic frequencies that propagate in both forward and backward directions with respect to the background magnetic field, **B**$_0$). Driven high frequency waves can play an important role in the particle acceleration, plasma heating, and turbulence generation [e.g., *Leamon et al.*, 1998;

*Matthaeus et al.*, 2008; *Breech et al.*, 2009; *Chandran*, 2010; *Hollweg et al.*, 2013]. As an example, generated high frequency waves have high dissipation rate, and they can also create a possibility of resonant wave-particle interactions using both channels: the normal cyclotron/gyro-resonance interactions and anomalous cyclotron interactions (this is when the positive ions interact with right-hand waves; see e.g., *Tsurutani and Lakhina*, 1997). Discussion of the specific effects that such driven waves might have in different laboratory, terrestrial, solar, or astrophysical plasma environment is beyond the scope of the current letter and will be discussed elsewhere.

**Acknowledgements**

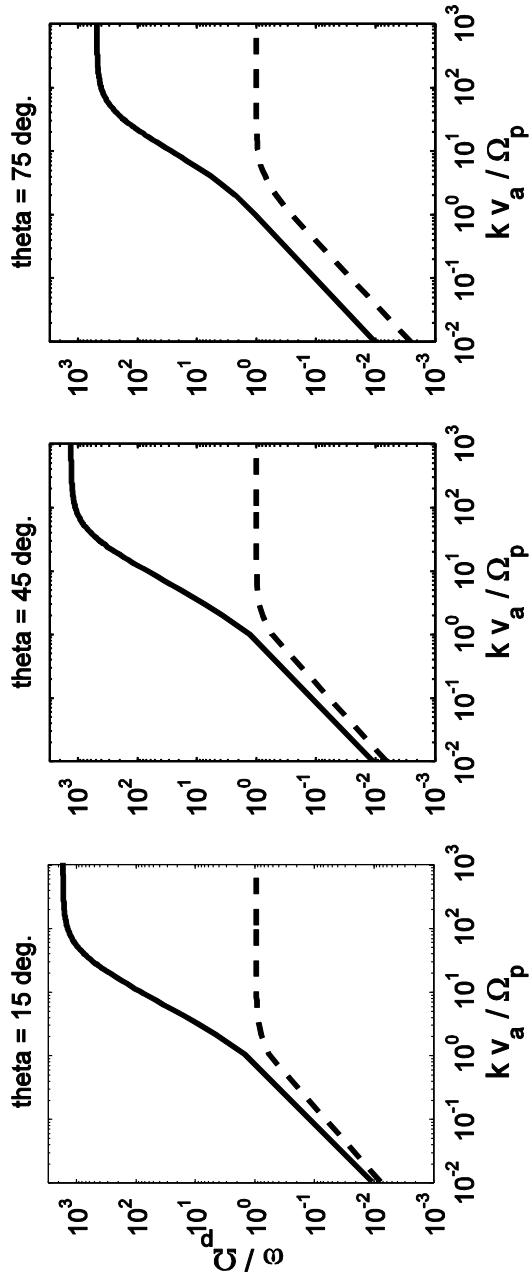

**Figure 1.** Modified Alfvén (dashed line) and fast magnetosonic wave (solid line) frequency plots. Three representative cases of the propagation angle between $k$ and $\mathbf{B}_0$ are chosen to show the separation of the frequency curves. For a given driver Alfvén wave, $kv_a/\Omega_p = const.$ vertical lines cross the frequency curves at the generated driven wave frequency values.